\documentclass[a4paper]{jpconf}
\usepackage{graphicx}
\usepackage{xspace}
\usepackage{relsize}

\def\C{\ensuremath{C}\xspace}
\def\P{\ensuremath{P}\xspace}
\def\T{\ensuremath{T}\xspace}
\def\CP{\ensuremath{CP}\xspace}
\def\CPT{\ensuremath{CPT}\xspace}

\def\Mz      {\ensuremath{M^0}\xspace}
\def\Mzb     {\ensuremath{\overline{M}^0}\xspace}

\def\babar{\mbox{\slshape B\kern-0.1em{\smaller A}\kern-0.1em
    B\kern-0.1em{\smaller A\kern-0.2em R}}\xspace}

\def\invfb   {\ensuremath{\mbox{\,fb}^{-1}}\xspace}
\def\invab   {\ensuremath{\mbox{\,ab}^{-1}}\xspace}

\begin{document}

\title{Experimental prospects for \C, \P, \T, \CP, and \CPT tests}

\author{Adrian Bevan}

\address{Queen Mary University of London, Mile End Road, London, E1 4NS, UK}

\ead{a.j.bevan@qmul.ac.uk}

\begin{abstract}
Discrete symmetry violation in the weak interaction is central to the Standard Model of particle physics,
however the origin of these violations is not well understood. Nor are we able to provide a satisfactory
explanation of the Universal dominance of matter over antimatter, an issue related to \CP violation.  
As a result study of discrete symmetry violation remains a topic of broad interest.  These proceedings discuss experimental prospects
of studying \C, \P, \T, \CP and \CPT symmetries in a number of contexts, including the 
use of triple product asymmetries and entangled neutral meson systems.
\end{abstract}

\section{Introduction}
The discrete symmetries parity (\P), charge conjugation (\C), time reversal which is also commonly
referred to as motion reversal (\T), and their 
combinations \CP and \CPT have been instrumental in probing a number of fundamental aspects of nature 
in the context of particle physics.  The strong and electromagnetic forces conserve these 
symmetries.  The weak interaction violates \P, \C, \CP, and \T, however the combination \CPT 
is found to be conserved by experimental observations made so far. \CP has been found to be 
violated in weak interactions involving quark transitions, and the quest for the complementary 
effect with neutrinos has started via combinations of data from all possible sources;
several new experiments have been proposed that may constrain or measure the \CP phase for
the neutrino sector.  Several tests of \CP and \CPT have been performed using charged leptons,
in particular $\tau$ decay studies of decays to $K_S\pi^\pm\nu$ and the mass difference between
$\tau^+$ and $\tau^-$, respectively~\cite{Bevan:2014iga}.
The underlying reasons behind the observed pattern of behaviour is not established, however
the observed phenomena have
been parameterised within the Standard Model of particle physics (SM).  

A common feature to testing symmetry violation is to identify a process $\psi$ 
and its symmetry conjugate $\psi^\prime$.  If a symmetry is conserved then the asymmetry
given by
\begin{eqnarray}
A=\frac{{\cal P}(\psi^\prime)-{\cal P}(\psi)}{{\cal P}(\psi^\prime)+{\cal P}(\psi)} \label{eq:asymmetry}
\end{eqnarray}
is zero as the probability for the process and its conjugate to occur are equal.  If
a measurement is performed where it is found that $A\neq 0$, then the symmetry is 
violated.  We can construct asymmetries using both particle transitions and ensembles of
particle transitions as reference processes.  The key is to identify reference states
where it is possible to physically define $\psi^\prime$ for the symmetry of interest.

The remainder of these proceedings discuss the use of triple product asymmetries to 
probe \P, \C and \CP violation followed by the use of entangled neutral meson systems to 
test \CP, \T and \CPT symmetries, before summarising.

\section{Triple product asymmetries}
\label{sec:tripleproducts}

\begin{figure}[!h]
\resizebox{0.99\textwidth}{!}{
  \includegraphics[width=6.5cm]{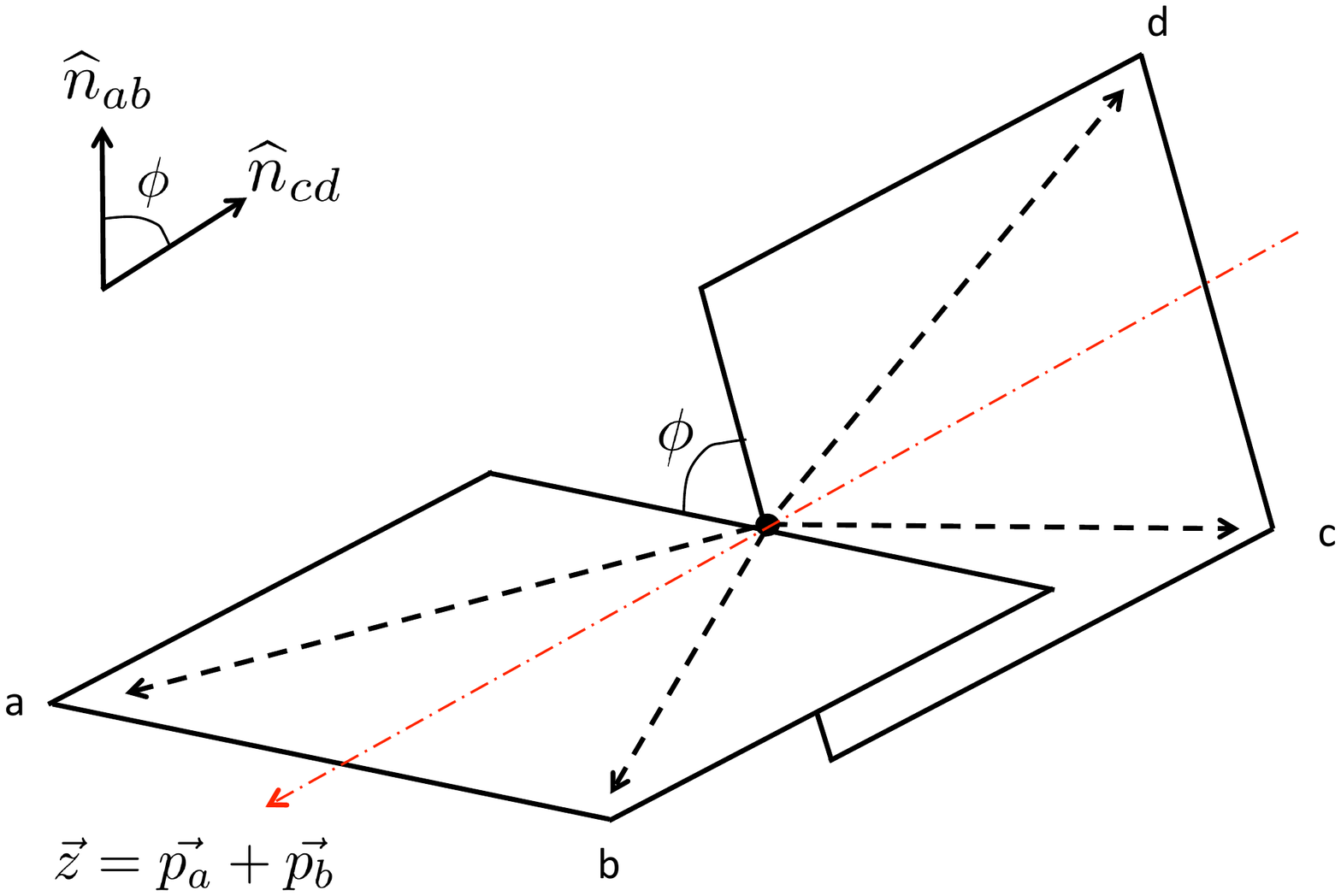}
  \includegraphics[width=6.5cm]{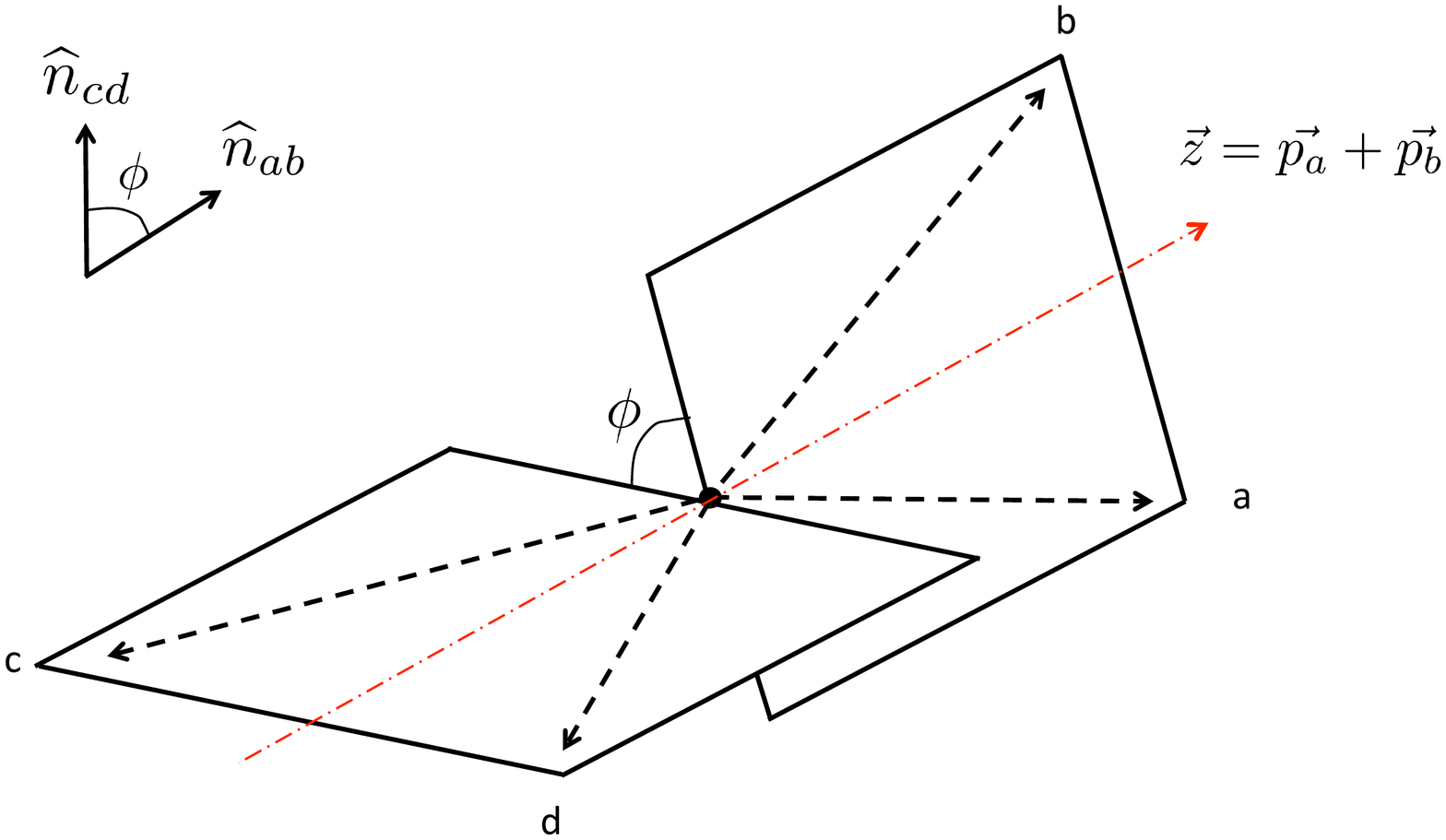}
}
\resizebox{0.99\textwidth}{!}{
  \includegraphics[width=6.5cm]{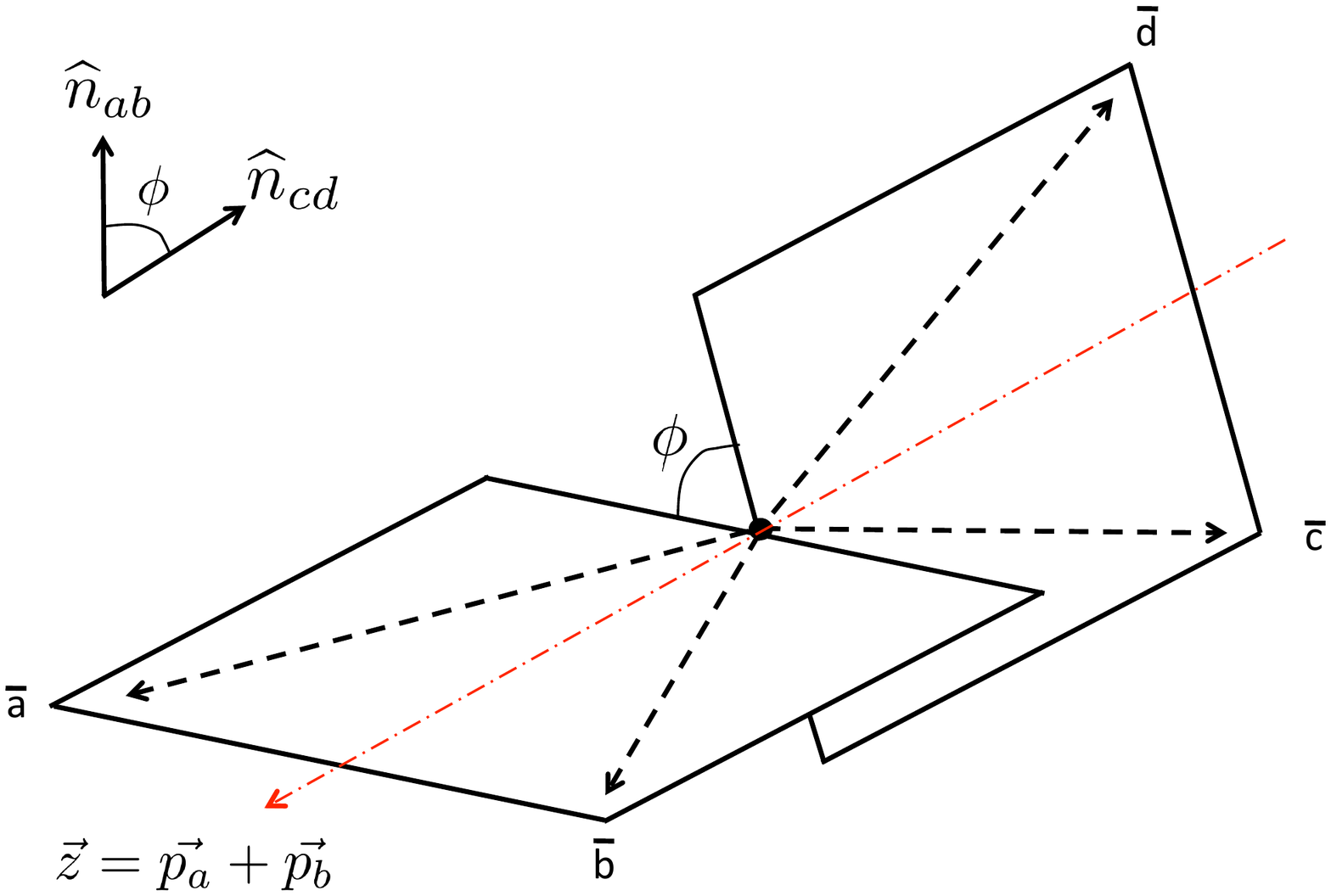}
  \includegraphics[width=6.5cm]{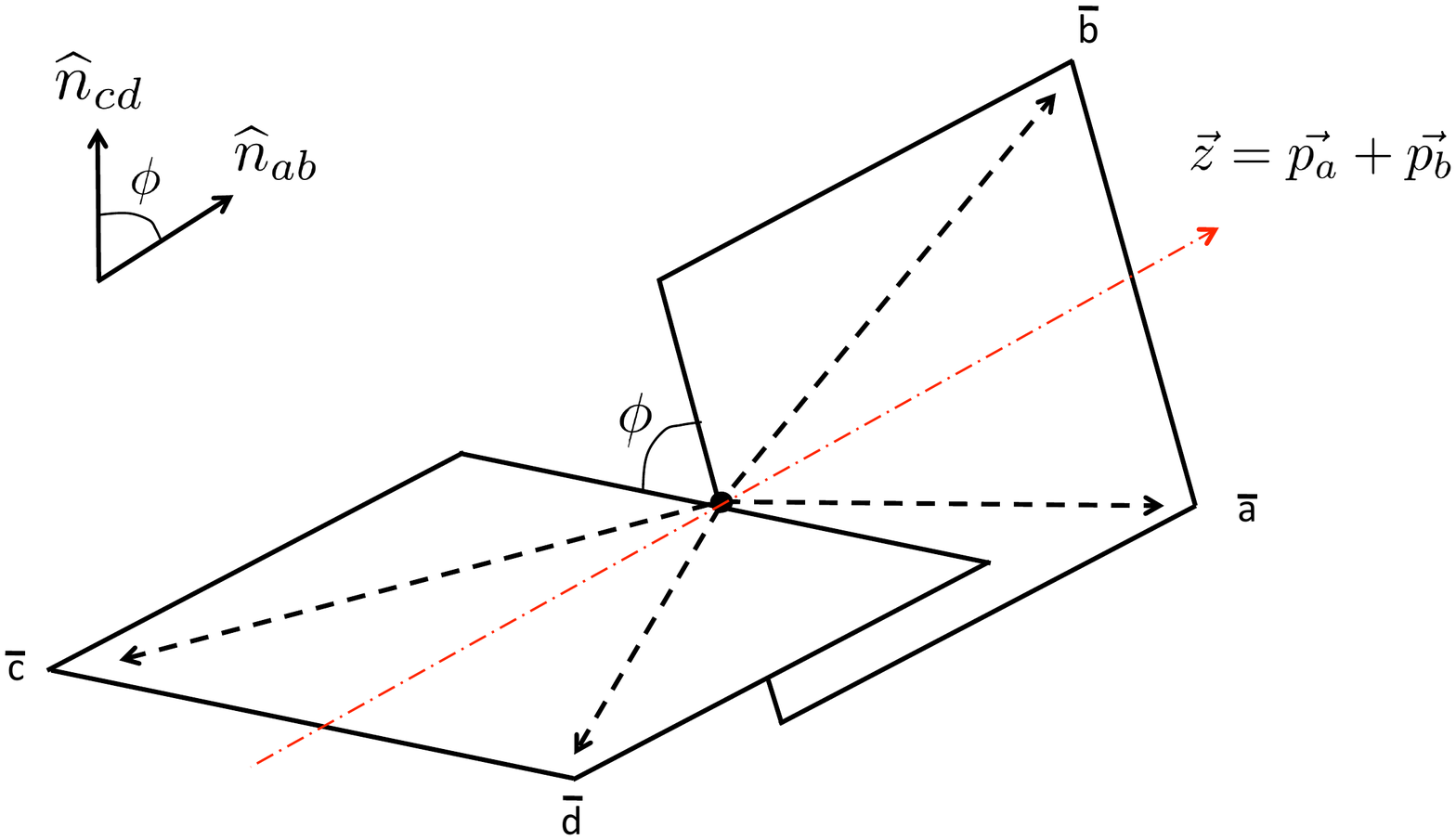}
}
\begin{minipage}[b]{0.99\textwidth}\caption{\label{fig:tripleproduct} The decay (top-left) $M\to abcd$ and the symmetry transformed 
decays under (top-right) \P, (bottom-left) \C, and (bottom-right) \CP.}
\end{minipage}
\end{figure}

Triple product asymmetries have been used to probe for \CP violation in a number of different systems, for example see~\cite{Gronau:2011cf,Bevan:2014nva} and references therein.  One can use the momenta (or spins) of final state particles in four body decays of some mother particle $M$ to a state $abcd$ to compute a scalar triple product $\vec{p}_c \cdot (\vec{p}_a \times \vec{p}_b)$.  The asymmetry of Eq.~(\ref{eq:asymmetry}) can be computed either integrating over positive ($+$) and negative ($-$) values of the triple product, or as a function of $\phi$, the angle between the decay planes defined by $ab$ and $cd$.  The scalar triple product is even under \C and odd under \P and \CP transformations.  
It is also odd under \T, however it is not possible to identify physical pairings of $\psi$ and $\psi^\prime$ for $M\to abcd$.  
Some of the literature chooses to invoke
\CPT in order to trivially relate \CP violation to a \T violation and call these \CP violating triple product asymmetry measurements \T violating.  
With such a mind set one ignores the full set of tests that can be performed and chooses the limiting case where 
all \CP violation signals might equally be called \T violation. 
For such an approach there is simply no point in making the distinction between triple product \CP tests and other \CP 
violation manifestations, hence no point in discussing \T violation.  
For some time a more general approach has been advocated in the literature, 
and in recent years experiment has gone beyond the limiting case of \CPT conservation 
for a number of measurements.  The invocation of \CPT conservation 
is no longer reasonable for a general discussion of discrete symmetry violations.
The latter part of these proceedings discuss tests of \CP, \T and \CPT symmetries, and these proceedings regard \CP violating triple products to be 
just that.  
It is worth noting that for entangled systems one could define $\psi$ and $\psi^\prime$ under a \T or \CPT transformation and hence study triple products using decays of $M \overline{M}\to (abcd)(efgh)$ as well as mixing 
\CP filters with the four body decays in analogy with the methodology outlined in Section~\ref{sec:entangledsystems}.  We do not discuss this possibility further here as there have not been any attempts to perform such a measurement thus far, however this is a further reason to ensure clarity of the language used when talking about symmetry violations\footnote{Several authors have pointed out that the usual definitions of discrete operators need 
to be modified in the context of \CPT violation.  For example see Ref.~\cite{Fidecaro:2013gsa}.}.  
Figure~\ref{fig:tripleproduct} illustrates the decay $M\to abcd$ and the corresponding \P, \C and \CP transformed processes.
The \T transformed process $\overline{abcd}\to \overline{M}$ never happens and is not shown.

By considering the possible conjugate pairings illustrated in the figure one can construct six distinct asymmetries from triple products.  These are~\cite{Bevan:2014nva}
\begin{eqnarray}
A_{\P} = \frac{\Gamma_{+} - \Gamma_{-}} {\Gamma_{+} + \Gamma_{-}}, &\,\,\,\,\,&
\overline{A}_{\P} = \frac{\overline\Gamma_{+} - \overline\Gamma_{-}} {\overline\Gamma_{+} + \overline\Gamma_{-}}.\label{eq:tp:parity}\\
A_{\C} = \frac{\overline\Gamma_{-} - \Gamma_{-}} {\overline\Gamma_{-} + \Gamma_{-}}, &\,\,\,\,\,&
\overline A_{\C} = \frac{\overline\Gamma_{+} - \Gamma_{+}} {\overline\Gamma_{+} + \Gamma_{+}}.\label{eq:tp:c}\\
A_{\CP} = \frac{\overline{\Gamma}_{+} - \Gamma_{-}} {\overline{\Gamma}_{+} + \Gamma_{-}}, &\,\,\,\,\,&
\overline{A}_{\CP} = \frac{\overline\Gamma_{-} - \Gamma_{+}} {\overline\Gamma_{-} + \Gamma_{+}},\label{eq:tp:cp}
\end{eqnarray}
where the quantities $\Gamma_{\pm}$ and $\overline{\Gamma}_{\pm}$ denote the rate for particle (anti-particle) decays with 
the sign of the triple product denoted by the subscript.  The subscript on the asymmetry $A$ denotes the symmetry used to 
construct the conjugate pair.    A further six asymmetries can be constructed by considering the two other symmetries acting
on $A$ and $\overline A$ in turn. These are given by~\cite{Bevan:2014nva}
\begin{eqnarray}
a_{\C}^{\P}&=& \frac{1}{2}\left( A_{\P} - \overline{A}_{\P}\right), \nonumber\\
a_{\CP}^{\P} &=& \frac{1}{2}\left( A_{\P} + \overline{A}_{\P}\right),\nonumber\\
a_{\P}^{\C} &=& \frac{1}{2}(A_{\C} - \overline{A}_{\C}),\nonumber\\
a_{\CP}^{\C} &=& \frac{1}{2}(A_{\C} + \overline{A}_{\C}),\nonumber\\
a^{\CP}_{\P} &=& \frac{1}{2}(A_{\CP} - \overline A_{\CP}),\nonumber\\
a^{\CP}_{\C} &=& \frac{1}{2}(A_{\CP} + \overline A_{\CP}).\label{eq:tp:derived}
\end{eqnarray}
Here these secondary asymmetries are denoted by a lower case $a$, where the superscript corresponds to the first symmetry
used for conjugate pairing and the subscript corresponds to the second one.  One can determine which symmetry is under scrutiny by multiplying the superscripts and subscripts together.  For example $a_{\C}^{\P}$ is a test of \CP, $a_{\CP}^{\P}$ a test of $\C\P^2 = \C$, and so on.
In general these asymmetries receive contributions
from all interactions, and the interest is in isolating effects that are dominated by weak interactions (i.e. theoretically
clean), or where that may not be possible, to isolate effects that can either be understood in the longer term, or
signify a non-trivial weak interaction effect even in the presence of pollution from strong force 
induced final state interactions (i.e. soft QCD or re-scattering).
A concrete example of this issue is discussed below in the context of the measurement of $\alpha_b$ from $\Lambda_b$ decays.
Hence interpretation of these asymmetries depends on the decay under study (and hence the model required to interpret data).

There are three categories of event that can be studied; for the first two all twelve quantities are non-trivial and distinct. For the third category the non-trivial asymmetries are degenerate.  These categories are defined by: (i) $M\neq \overline{M}$ and $abcd\neq \overline{abcd}$, (ii) $M\neq \overline{M}$ and $abcd = \overline{abcd}$, and (iii) $M = \overline{M}$ and $abcd = \overline{abcd}$.  
The asymmetry relevant for type (iii) decays is
\begin{eqnarray}
A_{\P, \CP} = \frac{\langle\Gamma\rangle_{+} - \langle\Gamma\rangle_{-} }{ \langle\Gamma\rangle_{+} + \langle\Gamma\rangle_{-}},\label{eq:tp:averagedratescp}
\end{eqnarray}
which can be constructed by considering \P and \CP using the rates measured.

\subsection{Meson decays}
\label{sec:tripleproducts:meson}

Triple product asymmetry measurements have been made in $K$, $D$, and $B$ systems.  Until recently only three of these were reported by experiments; $A_{\P}$, $\overline{A}_{\P}$, and $a_{\CP}^{\P}$.  The rare kaon decays $K_{L,S}\to \pi^+\pi^-e^+e^-$ are type (iii) processes and they have been studied experimentally by KTeV ($K_L$ only) and NA48 (both decays).    The PDG reports an asymmetry of $13.7\pm 0.5$\% for $K_{L}\to \pi^+\pi^-e^+e^-$~\cite{Agashe:2014kda}.
The results for the $K_{L}$ decay are in excellent agreement with theoretical prediction; a 14\% \CP asymmetry is observed~\cite{Heiliger:1993qt} in this decay.
The corresponding asymmetry found for $K_{S}\to \pi^+\pi^-e^+e^-$ is consistent with zero.
It was pointed out some time ago that $a_{\C}^{\P}$ could also be measured~\cite{Kang:2009iy} while studying triple product asymmetries relevant for charm studies.  
The full set of asymmetries has been measured by \babar for $D_{(s)}^\pm$ decays into $K_S K^\pm\pi^+\pi^-$ and for $D^0$ decays into $K^+K^-\pi^+\pi^-$~\cite{ckm2014}.  LHCb have recently measured the three asymmetries $A_{P}$, $\overline A_P$ and $a^P_C$ for these modes~\cite{Martinelli:2014nfa}.

The presence of soft QCD and re-scattering gives rise to the question, what is one testing with these decays.  It is well known that some asymmetries can be manifestly non-zero driven solely by strong phase differences and hence do not provide clear insight into weak dynamics.  The converse is also true for other asymmetries.  If one considers a simple model of two interfering amplitudes it is straightforward to show that half of the asymmetries defined above fall into the former category and half into the latter one.  Hence six of these quantities are interesting to measure from the perspective of the weak
interaction even in the limit that we are unable to control hadronic contributions.

\subsection{Other possible measurements}
\label{sec:tripleproducts:other}

It is possible to perform these measurements using baryon decays.  One possible advantage of such measurements is that the energy regime is increased relative to meson decays and so it may be easier to compute the expected contributions from hadronic uncertainties to estimate the underlying weak effects.  This is analogous to the effort underway for determining $\alpha$ from $\Lambda_b$ decays at the LHC~\cite{Aaij:2013oxa,Aad:2014iba} as a follow up to the corresponding study in $\Lambda$ decays performed by Cronin and Overseth~\cite{Cronin:1963zb}.  Similar measurements are proposed for $\Lambda_c$ decays~\cite{Bensalem:2002pz,Bensalem:2002ys,Kang:2010td}.  It is also possible to study top quark and tau lepton systems as outlined in~\cite{Bevan:2014nva}.

Since the advent of the LHC it has been possible to entertain the prospects of analysing data involving $Z$ or $H$ boson decay, or associated production of pairs of bosons ($ZH$, $WH$, $HH$, etc.) that subsequently decay into a four particle final state.  The SM predictions for $C$ and $CP$ effects are clear for all of these transitions - the associated asymmetries should be zero as non-zero violations arise from $W$ boson interaction with up and down type quark fields.  It is possible to estimate the precision with which the LHC can measure asymmetries for a number of modes based on the available Run 1 publications, and expectations through to the high luminosity LHC phase expected to start in the a decade from now.
One of the most straightforward measurements to make is that of a $Z$ or $H$ boson decay to four leptons; $\ell^+\ell^-\ell^{\prime +}\ell^{\prime -}$.  The final states where $\ell \neq \ell^\prime$ have the largest rate and no ambiguity in terms of combinatorics of the signal reconstruction.  The LHC has observed both $Z^0\to \mu^+\mu^- e^+e^-$~\cite{Aad:2014wra} and $H\to \mu^+\mu^- e^+e^-$~\cite{Aad:2012tfa,Chatrchyan:2012ufa}.
Expectations for the precision attainable for the triple product asymmetry measurement in these processes are given in
Table~\ref{tbl:boson}.  Percent-level precision could be achieved at the end of the high luminosity LHC (HL-LHC) assuming that trigger rates can be
maintained and systematic uncertainties can be controlled in reconstructing these final states.  Final states where
$\ell = \ell^\prime$ have rates a factor of two smaller than those for $\ell \neq \ell^\prime$, 
and may also suffer from a small dilution coming from ambiguities in reconstruction of the final states.  Hence one can
expect that the $4\ell$ modes, where $\ell = e, \mu, \tau$, have precisions at least $\sqrt{2}$ larger than the estimates shown in the table.

\begin{table}[!ht] \begin{center}
\caption {Estimated precisions on the asymmetry of Eq.~(\ref{eq:tp:averagedratescp}) for
$Z^0$ and $H$ decays to $\mu^+\mu^- e^+e^-$ for ATLAS and CMS at the LHC and HL-LHC.
These estimates are based on the published run 1 yields,
assuming that the asymmetry is zero.\label{tbl:boson}}
\renewcommand{\arraystretch}{1.2}
\begin{tabular}{c|cc} \hline\hline
Data sample                  & $Z^0\to \mu^+\mu^- e^+e^-$      & $H\to \mu^+\mu^- e^+e^-$ \\ \hline
Run 1 $(\sim 25 \invfb)$     & 0.12       & 0.38 \\
Run 2 $(\sim 125 \invfb)$    & 0.04       & 0.11 \\
Run 3 $(\sim 300 \invfb)$    & 0.03       & 0.07 \\
HL-LHC $(\sim 3000 \invfb)$  & 0.01       & 0.02 \\
\hline\hline
\end{tabular}\end{center}\end{table}

Another potential channel of interest is the associated production of a $H$ boson in conjunction with a $W$ or $Z$.
The use of triple product asymmetries to probe NP is $WH$ decays has been discussed in Ref.~\cite{Delaunay:2013npa}.
The latest experimental searches from the LHC are reported in~\cite{Chatrchyan:2013zna,Aad:2014xzb}.  It might be possible
to find evidence for these decays during Run 2.  If that were to happen one could start to measure triple product 
asymmetries for combinations of $H$ boson couplings with a $W$ or $Z$.  The proposed ILC is expected to 
accumulate large samples of $ZH$ bosons, and assuming the yields outlined in~\cite{tanabe:ipa} we would 
expect a statistical precision of 2.1 (3.7)\% and 2.5 (4.7)\% to be attainable on the asymmetry of
Eq.~(\ref{eq:tp:averagedratescp}) using inclusive (exclusive) $Z$ decays to lepton pairs
at 250 and 500 GeV, respectively.  The proposed Chinese Higgs Factory (CEPC) aims to integrate
a sample of about 1 million ZH decays~\cite{cepc}.  If built the proposed Future Circular $e^+e^-$ Collider (FCC-ee) 
would accumulate similar statistics.  Such a sample size could be used to archive a precision
about a factor of five times better than that of the ILC.

\section{Entangled states}
\label{sec:entangledsystems}

John Bell was the first person to understand the concept of quantum entanglement~\cite{EPR,Bell:1964kc} 
and in doing so solved an old problem in quantum mechanics.  This concept 
has been tested extensively in a number of systems, and is used in a practical sense today in the context of quantum information and to under-pin 
efforts to construct quantum computers.
This concept is also central to the study of discrete symmetry violation tests in neutral mesons performed at $K (\phi)$ and $B$ meson 
factories.  

Pairs of entangled neutral mesons $M=K,\, D,\, B_{s,d}$ can be prepared in decays of $\phi$, $\psi(3770)$, $\Upsilon(4S)$
mesons. If we assume the fundamentals of quantum mechanics
such as locality and coherence of the entangled wave function $\Psi$, then we may write
\begin{eqnarray}
\Psi &=& \frac{1}{\sqrt{2}} \left( M_1^0\overline{M}_2^0 - \overline{M}_1^0 M_2^0 \right),\\
     &=& \frac{1}{\sqrt{2}} \left( M_{1, +} M_{2, -} - M_{1, -} M_{2, +} \right).
\end{eqnarray}
Here the subscript $1$ and $2$ refers to the first and second meson, respectively and
$M^0$ is a particle, while $\overline{M}^0$ is the anti-particle for the flavour filter
basis set $\{M^0, \overline{M}^0\}$.  The corresponding \CP filter basis set is
$\{M_{+}, M_{-}\}$, where the subscript indicates the \CP eigenvalue of the filter.
One can probe \CP, \T, and \CPT using pairs of entangled mesons by considering the 15 non-trivial possible ways of constructing asymmetries.
These pairings are obtained by considering all possible combinations of decay for the first and second meson illustrated in 
Table~\ref{tbl:appendix_asymmetries:combinations}.

\begin{table}[!ht]
\caption{Possible combinations of decay for an entangled wave function.  The first meson decay type
is indicated in the first row, and that of the second meson is given by the first column.  \label{tbl:appendix_asymmetries:combinations}}
\begin{center}
\begin{tabular}{ccccc} \hline\hline
                 & $M^0$       & $\overline{M}^0$ & $M_+$ & $M_-$ \\ \hline
$M^0$            & $(+1, +1)$  & $(-1, +1)$  & $(\CP_{even}, +1)$  & $(\CP_{odd}, +1)$  \\
$\overline{M}^0$ & $(+1, -1)$  & $(-1, -1)$  & $(\CP_{even}, -1)$  & $(\CP_{odd}, -1)$ \\
$M_+$            & $(+1, \CP_{even})$  & $(-1,\CP_{even})$  & $(\CP_{even}, \CP_{even})$  & $(\CP_{odd}, \CP_{even})$ \\
$M_-$            & $(+1, \CP_{odd})$  & $(-1, \CP_{odd})$  & $(\CP_{even}, \CP_{odd})$  & $(\CP_{odd}, \CP_{odd})$ \\
\hline\hline
\end{tabular}
\end{center}
\end{table}

Traditionally experiments have studied entangled states in terms of flavour filter basis pairs only.  This permits two tests to be performed:
the comparison of $M^0 \to \overline {M}^0$ with the \CP and \T conjugate process
$\overline M^0 \to {M}^0$; the comparison of $M^0 \to M^0$ with 
the \CP and \CPT conjugate process $\overline M^0 \to \overline{M}^0$.  These two tests are typically characterised by reconstructing same sign and opposite
sign di-lepton final states in order to measure $A_{\CP, \T}$ and $A_{\CP, \CPT}$, 
respectively~\cite{Aubert:2002mn,Aubert:2003hd,Aubert:2004xga,Nakano:2005jb,Aubert:2006nf,Aubert:2007bp,Higuchi:2012kx}.  Some measurements have also been performed using
hadronically tagged flavour filters to complement those of semi-leptonic decays.  The study of hadronically tagged decays 
is motivated by the fact that the semi-leptonic decay results are dominated by systematic uncertainties at the 
$B$ Factories.
In particular 
measurements of $A_{\CP, \T}$ at the $B$ Factories mirror efforts by CPLEAR to measure \T violation via the so-called Kabir asymmetry method~\cite{Kabir:1970ts,CPLEAR}.  
The Tevatron and LHC experiments measure semi-leptonic asymmetries to perform these tests, see for example Ref.~\cite{Lenz:2011ti}.
It is interesting to note that these measurements serve as simultaneous tests of two symmetries at a time 
[c.f. the type (iii) triple product asymmetry above] and are constructed using only the flavour basis pairing.
Experimental results from the $B$ Factories on $A_{\CP, \T}$ and $A_{\CP, \CPT}$ measurements can be found in Chapter 17.5 of \cite{Bevan:2014iga}
and references therein.

\begin{figure}[!ht]
\begin{center}
\resizebox{0.95\textwidth}{!}{
  \includegraphics[width=12.5cm]{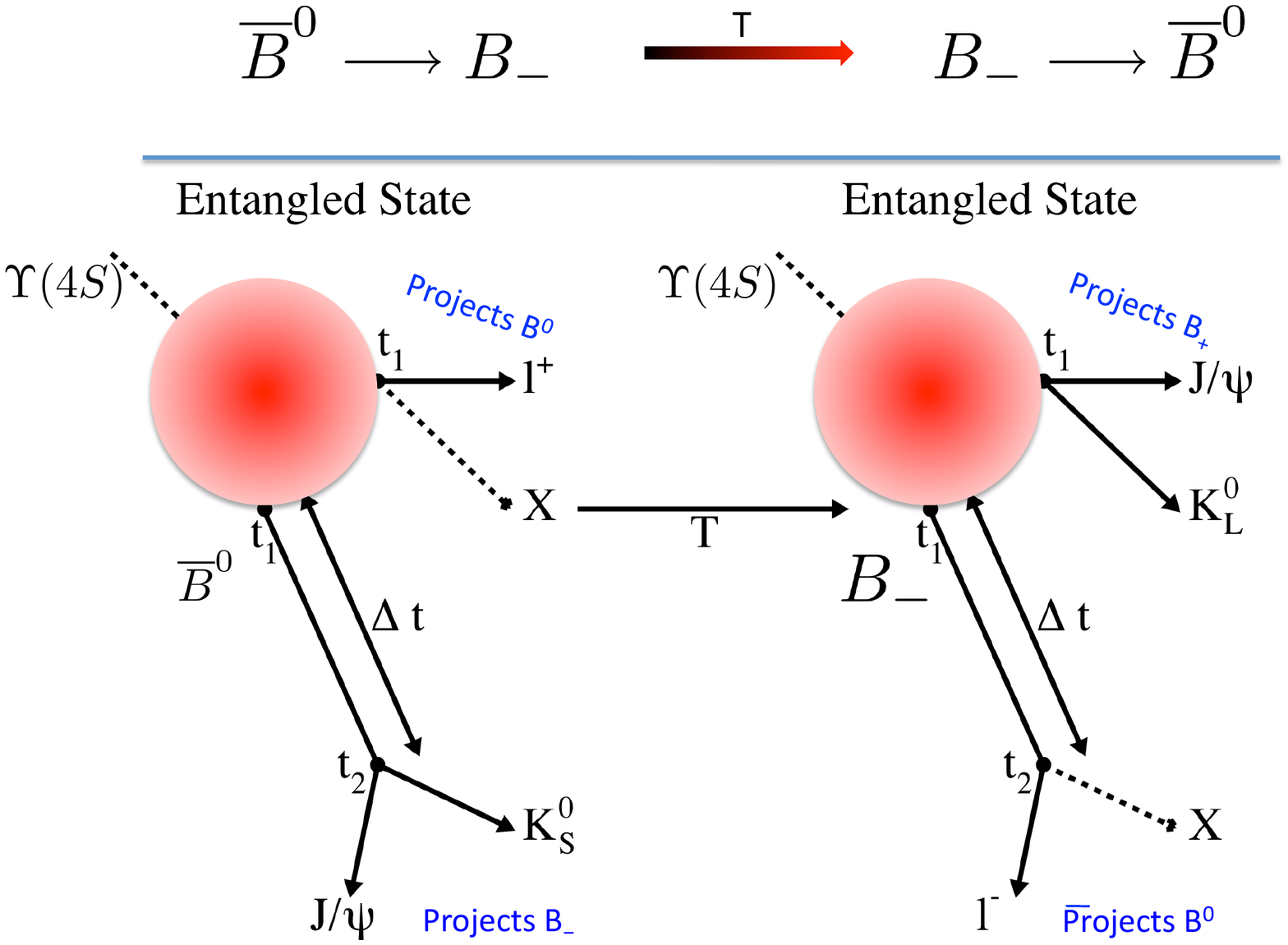}
}
\end{center}
\begin{minipage}[b]{0.99\textwidth}\caption{\label{fig:entangled} Illustration of the conjugate pairings used to test \T for a reference process of $\overline B^0\to B_-$ and the \T conjugate $B_-\to \overline B^0$.}
\end{minipage}
\end{figure}

Twelve asymmetries can be constructed as outlined by Bernabeu {\emph et al.} in Refs.~\cite{Banuls:1999aj,Banuls:2000ki,Alvarez:2006nk,Bernabeu:2012ab,Lees:2012kn}.  
The full set of reference and conjugate pairs used to test \CP, \T and \CPT are listed in Table~\ref{tbl:appendix_asymmetries:combinations:twelve}.
These are each tests of a single symmetry by virtue of invoking both flavour and \CP filter basis pairs reconstructed from the decays of the two mesons.  Figure~\ref{fig:entangled} illustrates this for the reference process $\overline B^0\to B_-$ under the symmetry \T.  The conjugate pairing to this under \T interchanges the initial and final states.  The $\overline B^0$ meson in the reference process is tagged by the decay of the other $B$ in the event being filtered by an $\ell^+ X$ decay and the $B_-$ filter is identified through the channel $J/\psi K_S$.  The \T conjugate of this scenario requires that the $B_-$ be tagged by the decay of the other $B$ in the event filtered via the channel $J/\psi K_L$, and the $\overline B^0$ is identified by the flavour filter channel $\ell^- X$.  It is not sufficient in this case to just consider the interchange of initial and final states, one also has to consider how to filter on the flavour and \CP components in order to correctly construct the desired asymmetry test.

\begin{table}[!ht]
\caption{The twelve pairings of reference and symmetry conjugated transitions used to study \CP, \T and \CPT
following the approach described in the text.} \label{tbl:appendix_asymmetries:combinations:twelve}
\begin{center}
\begin{tabular}{ccc} \hline\hline
Symmetry         & Reference transition & Conjugate transition \\ \hline
\CP              & $\Mzb\to M_-$ & $\Mz \to M_-$  \\
                 & $M_+\to \Mz$  & $M_+ \to \Mzb$ \\
                 & $\Mzb\to M_+$ & $\Mz\to M_+$   \\
                 & $M_-\to \Mz$  & $M_-\to \Mzb$  \\
\hline
\T               & $\Mzb\to M_-$ & $M_- \to \Mzb$ \\
                 & $M_+\to \Mz$  & $\Mz \to M_+$  \\
                 & $\Mzb\to M_+$ & $M_+\to \Mzb$  \\
                 & $M_-\to \Mz$  & $\Mz\to M_-$   \\
\hline
\CPT             & $\Mzb\to M_-$ & $M_- \to \Mz$  \\
                 & $M_+\to \Mz$  & $\Mzb \to M_+$ \\
                 & $\Mz\to M_-$  & $M_-\to \Mzb$  \\
                 & $M_+\to \Mzb$ & $\Mz\to M_+$   \\
\hline\hline
\end{tabular}
\end{center}
\end{table}

The \babar experiment has performed the measurements proposed by Bernabeu {\emph et al.}, the results of which can be found in~\cite{Lees:2012kn}.
\babar find \T violation occurs with a significance, assuming Gaussian errors, of $14\sigma$.  This balances the level of \CP violation observed, and \CPT is found to
be conserved.  The asymmetries are measured as a function of the time difference $\Delta t = t_2-t_1$, which is positive by definition and not
to be confused with the signed difference in proper time used in usual time-dependent \CP analysis approach.  The time-dependent rates
are given by
\begin{eqnarray}
g^\pm_{\alpha,\beta} (\Delta t)\propto e^{-\Gamma \Delta t}\left[ 1 + C^\pm_{\alpha,\beta}\cos(\Delta m \Delta t) + S^\pm_{\alpha,\beta}\sin (\Delta m \Delta t)\right],
\end{eqnarray}
where $\Gamma$ is the width of the $B$, $\Delta m$ is the mixing frequency between $B^0$ and $\overline B^0$ mesons, $\alpha$ and $\beta$ are the flavour and \CP filters and $\pm$ indicates if the flavour filter occurs before the \CP filter ($+$) or not ($-$).  One can relate the $S$ and $C$ parameters to the real and imaginary parts of $\lambda=(q/p)(\overline{A}/A)$, where $q$ and $p$ are mixing parameters and $\overline{A}/A$ is the ratio of amplitudes for antiparticle and particle decays to 
\CP conjugate final states. This means that these measurements can be used to constrain the phase in the CKM matrix.  For example
the \T asymmetry can be written as
\begin{eqnarray}
A_T \simeq \frac{\Delta C^\pm}{2}\cos(\Delta m \Delta t) + \frac{\Delta S^\pm}{2}\sin(\Delta m \Delta t),
\end{eqnarray}
where in the SM one would expect $\Delta C^\pm = 0$ and $\Delta S^\pm= \mp \sin 2\beta$ for the \CP filters $J/\psi K_{S,L}$.

It has been noted that one has to take care in defining the operators used to test symmetries when \CPT is relaxed. A potential consequence of
this is a second interpretation of the \babar results; it is possible that those results are a sign of \CPT violation masquerading as fake \T violation~\cite{Fidecaro:2013gsa}.  The \babar interpretation of the data is the more conservative approach, and it will be very interesting to see the results of complementary measurements
made by Belle II in the fullness of time in case this might shed light on the interpretational ambiguity.

It has been noted that one can extend the set of \CP filter basis pairs from $\{J/\psi K_S, J/\psi K_L\}$ to other final states in order to broaden the possible measurements that can be made~\cite{Bevan:2013wwa,Dadisman:2014mya}.  In addition to studying $c\overline{c}s$ \CP filter transitions where the reconstructed kaon is either used to select a $K_S$ or $K_S$ filter state.  
One can use decays into $VV$ final states to filter out \CP even an odd amplitudes via a time-dependent angular analysis as noted in~\cite{Bevan:2013wwa}.  These measurements are yet to be performed, however it is expected that Belle II would accumulate sufficient statistics in 50\invab to observe \T violation using \CP basis pairs from $(\eta^{(\prime)},\omega) K_{S,L}$, $\phi K^*$ and $D^*D^*$ decays.

The remaining asymmetry that can be constructed with entangled states is done so by considering only the \CP filter basis pairing. This enables one
to compare $M_{+}\to M_{-}$ with $M_{-}\to M_{+}$ and in doing so perform a test of \T and \CPT; the rate difference between this pair of processes is $A_{\T, \CPT}$.  
Experimental prospects for such a measurement are bleak at the current time. \babar and Belle should have of the order of 1 event, and Belle II is 
expected to accumulate of the order of 20 events for such a measurement.  A third generation $B$ factory may be able to start measuring $A_{\T, \CPT}$.

As remarked in Section~\ref{sec:tripleproducts} it is possible to also test \C and \P using entangled meson systems if one studies decays into double-tagged pairs of four body final states, or other combinations of final state using flavour or \CP filter decays.  Hence it is possible to probe the full set of weak interaction behaviour under \C, \P, \CP, \T and \CPT using the same sample of data to fully over-constrain the consistency of results with expectations of the weak interaction.  

\section{Summary}

The weak interaction is well known to violate \P, \C, \T and \CP.  These effects are accommodated in the SM Lagrangian, however 
a deep understanding of the origin of these phenomena eludes us.  The potential to learn something new requires 
a next generation of innovative and exhaustive measurements in the hope that some may go beyond our current expectations
and hint at a deeper truth.  The combined symmetry of \CPT has important ramifications, and as a consequence we require
\CPT to be conserved if Lorentz symmetry is respected in nature.  
While one expects to find that this symmetry is conserved, at some point we believe that \CPT violation may be 
manifest (for example in theories of quantum gravity).  These measurements are important to make
as the implications, should we find \CPT violation, would be far reaching for modern physics.  
The use of triple product asymmetries in testing symmetry violation for different final states
may yield a deeper insight into the weak interaction.  Likewise the use of entangled meson systems
at experiments will continue to enhance our understanding of violations of these discrete symmetries.  By systematically
probing a broad range of observables spanning quarks, leptons and bosons one hopes that the data may point to a more
complete understanding of nature.

\end{document}